\documentclass[twocolumn,nobibnotes,showpacs,showpacs,preprintnumbers,amsmath,amssymb,superscriptaddress]{revtex4}
\usepackage{graphicx}
\usepackage{dcolumn}
\usepackage{bm}
\usepackage{lineno}
\begin{document}

\preprint{}

\title{Persistence, extinction and spatio-temporal synchronization of
SIRS cellular automata models}
\author{Quan-Xing Liu}
\email{liuqx315@sina.com}
\affiliation{Department of Mathematics, North University of China,\\
Taiyuan, Shan'xi 030051, People's Republic of China
}%
\author{Rong-Hua Wang}%
\email{rhwang624@sina.com}
\affiliation{Department of Mathematics, North University of China,\\
Taiyuan, Shan'xi 030051, People's Republic of China
}%

\author{Jin Zhen}
\altaffiliation[]{Corresponding author} \email{jinzhn@263.net}
\affiliation{Department of Mathematics, North University of China,\\
Taiyuan, Shan'xi 030051, People's Republic of China
}%

\date{\today}

\begin{abstract}
Spatially explicit models have been widely used in today's
mathematical ecology and epidemiology to study persistence and
extinction of populations as well as their spatial patterns. Here we
extend the earlier work--static dispersal between neighbouring
individuals to mobility of individuals as well as multi-patches
environment. As is commonly found, the basic reproductive ratio is
maximized for the evolutionary stable strategy (ESS) on diseases'
persistence in mean-field theory. This has important implications,
as it implies that for a wide range of parameters that infection
rate will tend maximum. This is opposite with present results
obtained in spatial explicit models that infection rate is limited
by upper bound. We observe the emergence of trade-offs of extinction
and persistence on the parameters of the infection period and
infection rate and show the extinction time having a linear
relationship with respect to system size. We further find that the
higher mobility can pronouncedly promote the persistence of spread
of epidemics, i.e., the phase transition occurs from extinction
domain to persistence domain, and the spirals' wavelength increases
as the mobility increasing and ultimately, it will saturate at a
certain value. Furthermore, for multi-patches case, we find that the
lower coupling strength leads to anti-phase oscillation of infected
fraction, while higher coupling strength corresponds to in-phase
oscillation.
\end{abstract}

\pacs{87.23.Cc, 
05.40.-a, 
82.40.Ck
}

\keywords{Epidemiology; Phase transitions;
Emergence of trade-offs; In-phase}
\maketitle

\section{Introduction}

In recent years, there has been an increased awareness of the
threats posed by newly emerging and high-profile infections disease,
such as SARS, the H5N1 strain of avian influenza, HIV, Ebola,
Whooping cough, Dengue fever and the spread of influenza. These
diseases exhibit large scale spatial contagions and long-term
spatial
patterns~\citep{Ballegooijen12282004,CecileViboud04212006,Daihai2003,
Nguyen2008,Yingcun2004,Earn1998,Heino1997,Bolker1995,DavidJ20000}.

At present, there is a great deal of interest in the role of spatial
structure in both ecology and epidemics. All systems are to some
degree spatially extended; however, the classical theory of the
dynamics of epidemics ignores spatial effects with its assumption of
homogeneous mixing. It is likely that local processes will play an
important role in the majority of infectious diseases' interactions
particularly when an infection occurs through the direct contact of
infected and susceptible individuals~\citep{Frank1991,Rand1995}. As
a consequence, there has been some attempt to take explicitly
spatial structure into account when examining the spreading of
parasites~\citep{Keeling1999,Keeling06152004,Ballegooijen12282004,
Fuks2006,liu:031110,Sasaki1999,Sasaki2007}. For example, recent
models show that interactions between local populations  may
generate complex spatial patterns by
dispersal~\citep{Fuks2006,liu:031110,Ballegooijen12282004,
CecileViboud04212006,Grenfell2001}. In modern societies, individuals
can easily travel over a wide range of spatial scales. The
interconnections of areas and populations through various means of
transport have important effects on the geographical spread of
epidemics~\citep{CecileViboud04212006}. In particular, the spatial
structure and the different levels of the diffusion and transport
processes are responsible for the heterogeneity, that is an erratic
outbreak patterns observed in the worldwide propagation and
persistence of
diseases~\citep{VittoriaColizza02142006,Ballegooijen12282004,
Daihai2003,Stone2007,Earn1998}, recently documented for
synchronization and waves~\citep{Daihai2003,Stone2007,Grenfell2001,
PhysRevLett.86.2909,CecileViboud04212006,Bolker1995}, and
large-scale spatial patterns (e.g. spiral
waves~\cite{Ballegooijen12282004,liu:031110,Jeltsch1997} and Turing
patterns~\cite{1742-5468-2007-11-P11011,1742-5468-2007-05-P05002})
in measles, dengue fever, SARS, and influenza. In order to describe
such a complex phenomenon and develop powerful numerical forecasting
tools, different levels of description are possible, ranging from a
simple global mean-field to detailed individual-based simulation
(see Refs.~\cite{1742-5468-2007-09-L09001,Keeling1999,
PhysRevLett.86.3200,Lloyd05182001,Ferguson2003,Stephen2003,DeAngelis1992}
and references therein), cellular
automata~\citep{0305-4470-25-9-018,0305-4470-26-15-020,
PhysRevE.50.4531,Fuks2006,liu:031110,Ballegooijen12282004,Jeltsch1997,durrett:677},
coupled map lattices~\citep{Sherratt1997}, etc.

 Dispersing individuals may react in a complex
manner to local ecological and epidemiological
situations~\citep{CecileViboud04212006,
Yingcun2004,Lima1996,Schippers1996,Jeltsch1997}, but rare
long-distance dispersal events may be important in
nature~\citep{FRAdler1993,Allen1993, Nguyen2008,Jeltsch1997}. The
modeling methods that summarize dispersal in a diffusion coefficient
in the classical theory--partial differential equations--cannot give
insights into the importance of rare events~\citep{Sasaki1999}. For
humans and other social animals in which hosts are distributed in
heterogeneous patches on a large scale, there are two critical sides
to transmit. The first is local transmission among individuals
within patches or communities (cites, towns, and villages). The
second is the transmission between patches. With this point of view,
there is a clear conceptual link between ecological and
epidemiological theory; a body of recent works have focused on the
analogies of the spatial dynamics between
 infectious disease and ecological
metapopulations~\citep{Earn1998,Yingcun2004,
Grenfell1997,Bolker1999}. The twofold issues are to dissect how
infection processes at the local scale determine spatio-temporal
patterns of epidemics and understand how these patterns are affected
by the spatial spread between neighborhoods or  patches, for
instance, the epidemic wavefronts observed  in the spatio-temporal
spread of the Black Death in Europe from 1347 to
1350~\citep{Murray1993,Grenfell2001} and West Nile
virus~\citep{MarkLewis2006}.

Generally speaking, several generic questions can be asked when
studying the dynamics of epidemic spreading. One is the explanation
of the possible oscillations~\footnote{ The sustained oscillations
include the coherence resonance phenomenon in the epidemic
models~\cite{Kuske2007,JonathanDushoff11302004,Simoes2008,Alonso2006}.}
with respect to the temporal evolution of the densities of infection
waves in homogeneous or heterogeneous realistic situation, as well
as the spatial resonance problem. Another one concerns the study of
the possibility that
 a disease spreading eventually arrives at a steady states and the time
 arrived~\citep{1742-5468-2007-09-L09001,Gautreau2007}.
In addition, our model can also investigate what enables some
species to persist while others become extinct. This question has
shaped the history of research on population dynamics and remains a
central issue for ecologists and epidemiologists~\citep{Earn1998}.

Although the spatial structure has been developed for
susceptible-infected-resistant (SIRS) models in previous studies,
the modulation of such effects in combination with individuals'
mobility is also unknown. We aim to bridge this gap in this paper.
Specifically, we aim to elucidate the phase space on the extinction
and persistence and contrast them with the prediction obtained
mean-field, as well as the effect of mobility on persistence
criteria. We also examine how the spatial pattern depend on the
mobility and coupling strength within multi-patches case.

\section{Model and methods}

\subsection{Cellular automata model}

In a recent report, van Ballegooijen and Boerlijst (vBB) showed that
there were three types of spatial pattern in a spatial
susceptible-infected-resistant model for disease dynamics by using a
grid-structured contact network, also called cellular automata
model~\cite{Ballegooijen12282004}. They were localized disease
outbreaks with self-limiting in size, turbulent waves, and stable
spiral waves respectively. Furthermore, they predicted that there
existed a trade-off between the parameters of infection period and
infection rate, which emerges from the evolutionary dynamics of the
system on the stable spiral waves, and referred to this relationship
as \textit{emergent trade-off}. These results give a guide to
understand the spatial features of epidemic such as wave speed and
wavelength, as well as the persistence and extinction relationship
depending on the parameters. To model the spatio-temporal phenomena,
a feasible approach is the combination of grid-based models and
movements of individuals. In this paper we take the framework
proposed by vBB as a starting point in assessing the phase
transitions between the persistence and extinction rather than
emergence of the stable spiral waves.

We use a spatial susceptible-infected-resistant model. The
population with $N$ individuals is categorized according to its
infection status: susceptibility (S), infectious (I), or resistant
(R)~\citep{Anderson1991,Ballegooijen12282004,Daihai2003,Stone2007}.
Within a subpopulation, the dynamics for the local populations obey
a basic reaction scheme conserving the number of population, which
has been studied both in physics and mathematical epidemiology,
namely the stochastic infection dynamics process identified by the
following set of reactions~\citep{Ballegooijen12282004}
\begin{subequations}\label{eq:reaction}
\begin{equation}\label{eq:reaction1}
S+I \stackrel{\beta}{\longrightarrow} I+I,
\end{equation}
\begin{equation}\label{eq:reaction2}
I \stackrel{\tau_{I}}{\longrightarrow} R,
\end{equation}
\begin{equation}\label{eq:reaction3}
R \stackrel{\tau_{R}}{\longrightarrow} S.
\end{equation}
\end{subequations}
The first reaction (Eq.~\eqref{eq:reaction1}) reflects the fact that
an infected host (I) can infect susceptible (S) neighbors with
infection rate $\beta$; the second reaction
(Eq.~\eqref{eq:reaction2}) indicates the acquisition of resistance
that hosts are infectious for a fixed period $\tau_{I}$, after which
they become resistant (R); and the third reaction
(Eq.~\eqref{eq:reaction3}) indicates the loss of resistance that
after a fixed period $\tau_{R}$, resistant hosts once again become
susceptible. In the sense of individual/local-level model, infected
hosts can infect adjacent susceptible hosts at infection rate
$\beta$. Here, the infection rate is not equal the probability of
infection, see~\citep{Ballegooijen12282004,Brauer2000,Yingcun2004}.
In the short-range case, the infectious neighborhood consists of
two/eight direct nearest-neighbors (NN) in a 1D/2D square lattices.
In addition, some recent investigations have included long-range
processes in one- and two-dimensional lattice models, reporting
quite different results on the occurrence of self-sustained
oscillation, extinction and
persistence~\citep{PhysRevE.63.061907,PhysRevE.63.056119,antoniazzi:040601,
tessone:224101,PhysRevE.62.8409,rauch:020903,1742-5468-2005-09-P09002,
Rozenfeld2004,Rozenfeld2001}, as well as increasing coupling between
subpopulation (subcommunity) in spatially structured environment can
lead to population
outbreaks~\citep{Petrovskii2001,Yingcun2004,CecileViboud04212006}.
Thus, an understanding of the joint effect of short- and long-range
interaction on the individuals (including the mobility called as the
diffusion, hopping, or stirring, hereafter, we refer as mobility) is
desirable from ecological and statistical physical point of
view~\citep{mobilia:040903}. Of course, in present paper, we also
further study the outbreak of disease inspired stochastic lattice
epidemic model with a nearest-neighbor interaction.

For ease of comparison, we follow the vBB's model by considering a
regular network of sites, each of which contains one of a single
susceptible individual (S), an infected individual (I) and resistant
(R). The susceptibles are infected by contact with an infected host
at a rate $\beta$, and the transmission can only occur locally.
Every time-step $\Delta t$, cells' state change according to the
rules in earlier studies~\citep{Ballegooijen12282004,Daihai2003},
which are illustrated in the corresponding Eq.~\eqref{eq:reaction}.

\subsection{Rate equations}

The deterministic rate (or mean-field) equations  describe the
temporal evolution of the stochastic lattice system, defined by the
reactions~\eqref{eq:reaction}, in a mean-field context, i.e. they
neglect all spatial correlations. They may be seen as a
deterministic description (for example emerging in the limit of
large system sizes) of systems without spatial structure. The study
of the rate equation is the ground on which the analysis of the
epidemics spreading and outbreak. In particular, the properties of
the rate equations are extremely useful for the epidemics
persistence by estimated values for the classical expression of
$\mathcal{R}_0$--the basic reproduction number. $\mathcal{R}_0$ is
defined as the number of secondary infections caused by a single
infective during its infectiousness period in an entirely
susceptible population~\citep{Anderson1991,Murray1993}. For the
standard SIRS mean-field approximation, also referred in
reaction~\eqref{eq:reaction}, are as follows:
\begin{subequations}\label{eq:mf}
\begin{equation}\label{eq:mfa}
\frac{dS}{dt}=-\beta SI+\tau_{R} R,
\end{equation}
\begin{equation}\label{eq:reaction2}
\frac{dI}{dt}=\beta SI - \tau_{I} I,
\end{equation}
\end{subequations}
where $S$ and $I$ denote the number of susceptibles and infected,
respectively. The number of recovered individual $R$ is obtained by
conservation of the entire population, i.e., $R(t)=N-S(t)-I(t)$. The
key quality describing the infection is the basic reproduction
number $\mathcal{R}_0=\beta/\tau_{I}$. If $\mathcal{R}_0>1$ and the
initial relative number of susceptibles are greater than a critical
value $S_{c}=1/\mathcal{R}_0$, an epidemic develops ($dI/dt>0$). As
the number of infected individuals increases, the number of
susceptibles $S$ decreases, and thus the number of contacts of
infected individuals with susceptibles decreases until $S=S_{c}$,
when the epidemic reaches its maximum and subsequently decays. This
processes alternative again with the time going if
$\mathcal{R}_0>1$. Fadeout is most likely to occur in the period
immediately following a major epidemic (the so-called inter-epidemic
trough). An important observation is that deterministic models
cannot reproduce such effects: the ability of populations to recover
from very low levels is a well-known weakness of many deterministic
models~\cite{Bolker1995,Lloyd2004}.

\section{Simulation and Results}

To begin with, we revisit some of the findings by vBB
\cite{Ballegooijen12282004}. Their results then serve as a reference
point for illustrating the difference of our elaborated approach
from their results. This is justified by the fact that vBB only
provide the spatial pattern resulting from static dispersal among
individuals within single patch, as well as emergence of trade-offs
between spiral waves and turbulent waves. Our crucial measures are
the phase transition between the global extinction and persistence
of the epidemic, and the effect from the movements of individuals on
it. In these simulations we set there is no migration among patches,
i.e., coupling parameter $LR=0$, in which our simulations faithfully
repeat the general findings by VBB~\cite{Ballegooijen12282004} that
three types of spatial patterns emerge (see Fig.~\ref{fig3}) but
find the differences when the mobility of individuals is turned on.

\subsection{Phase transitions of persistence and extinction}\label{section:short}

In this subsection, we analyze the relationship of infection rate
and infection periodic on phase transition between global
persistence and extinction within single patch. In the final
subsection, we will introduce the multipatches into the spatial
structure model by the coupling strength between the two patches.

When the individuals are statical in the local site, then the
infection only happens in the nearest neighbors, this form referred
to as \textit{local interaction}. Specially, we can study the
dynamics of such spatially extended model in one- or two-dimensional
space by defining appropriate microscopic rules.
Equation~\eqref{eq:reaction} shows the explicit transition rates of
cells' state in space, where the susceptibles infected by infectious
is isotropic. Grid size used in the simulations is $200\times200$
cells and time-step equals to 0.01. Larger grid sizes do not change
the evolutionary dynamics (see the computer code for the CA model at
http://7y.nuc.edu.cn/jinzhen/English1.aspx).

The persistence phase space is that the epidemics will persist when
the parameters lie in this domain. Here, we refer to
$(\tau_{I},\beta)$-parameters space. Their values are determined by
simulating $200\times200$ cell lattices with ten independent
runnings. Each lattice is initially set with random occupied by 100
infected individuals. For each obtained data, we fix the infection
period $\tau_{I}$, and chose an initial infection rate $\beta$.
Then, increased $\beta$ by small step (we use $\Delta \beta=0.05$)
and 10 thousand time units are simulated with that infection rate
value to determine the critical value of the phase transition. Here,
we find that 10 thousand time units is a long-term run for
persistence on this system. However, the typical waiting time $T(N)$
until extinction occurs is generally very long when the system size
$N$ is large. This suggests to consider the dependence of the
waiting time $T(N)$ on $N$. Quantitatively, we discriminate between
persistence and extinction by using the concept of extensively,
adapted from statistical
physics~\citep{reichenbach:051907,reichenbach2007nature}. If the
epidemic is not extinct at that time, the pair parameters' values
are assumed to be within the persistence space. In the present
model, we find that the epidemic will be persistence when parameter
$\beta$ is between a minimal critical infection rate $\beta_{min}$
and a maximal critical infection rate $\beta_{max}$. It is also
worth noticing that the definitions of persistence and extinction in
the presence of absorbing
states are intimately related to the concept of long-term. 

The threshold result for the deterministic model can be used to
identify two domains in the space of essential parameters
$\mathcal{R}_0$, namely $\mathcal{R}_0\leq0$ and $\mathcal{R}_0>0$,
where the model solutions behave in qualitatively different ways. We
endeavor to describe the spatially stochastic model in a similar
way, by identifying domains in the space of its essential parameters
where model solutions behave in qualitatively different ways. We
perform extensive computer simulations of the
system~\eqref{eq:reaction} on the phase transition, which describes
the persistence and extinction with respect to the
$(\tau_{I},\beta)$-parameters space shown in Fig.~\ref{extinction}
which also demonstrates that there exists a trade-off between
infection period and infection rate. In order to compare the classic
mean-field approximation of non-spatial theory $\mathcal{R}_0$ and
the spatial structure of individuals on the phase transition, we use
the power laws to fit the simulation data. Moreover, previous
studies show that the infection rate ($\beta$) and the virulence
($\alpha$) also have a relation of power laws on the parameters of
the infection rate and virulence~\citep{Boots1999,Boots2003}. Yet,
this relationship can be expressed generally with $f(x)=ax^b+c$
formula, as shown in Fig.~\ref{extinction}, in which the red curve
and blue curve are fitted by this function with 95\% confidence
intervals on the parameter estimates, respectively. As could be
expected, there exist
 minimal and maximal critical values of the infection rate $\beta$
on the phase transition. The open circles ($\circ$) and faced
circles ($\bullet$) indicate the simulated results of the minimal
critical infection rate $\beta_{min}$ and maximal critical infection
rate $\beta_{max}$, respectively. By comparison between fitted
curves and simulations, one can see there is a nonlinear trade-off,
$\beta(\tau_{I})$ is a monotonically decreasing function of
$\tau_{I}$ and bounded by a positive constant. The dashed line in
Fig.~\ref{extinction} show the prediction on the phase space from
mean-field theory on $\mathcal{R}_0$. From the classical expression
of $\mathcal{R}_0$, we know that the diseases will die out if
$\mathcal{R}_0\leq0$. This has important implications, as it implies
that for a wide range of $(\beta,\tau_{I})$ space there are two
parameter domains in the $(\beta,\tau_{I})$ space. One is domain of
extinction and the other is  domain of persistence. This contrast
with the result obtained in spatially explicit model where the
parameter domain is divided into three: domains (I) disease free;
domain (II) endemic (see Fig.~\ref{extinction}). The numerical
results clearly indicate (see Fig.~\ref{extinction}, Inset) the
validity of the prediction for the spatial thresholds to control
invasion of parasites. In the $\log$-$\log$ plot (Inset of
Fig.~\ref{extinction}), the dashed line and power laws ($c=0$) is
straight line of slope $-1$ shown on the graph. Then, the same where
 $\log(f(x)-c)$ is plotted against $\log(x)$. One can see that a nonzero
 $c$ (nonvanishing infection rate) is better straight line. In this
case, the nonzero $c$ corresponds to the phase boundaries for large
$\tau_{I}$. The best-fit line on a $\log$-$\log$ scale has a slope
of -0.7065, indicating a linear dependence of $\log\tau_{I}$ on
$\log\beta$. An understanding of this scaling behavior is lacking.

\begin{figure}[htp]
\centering (a)\includegraphics[width=4cm]{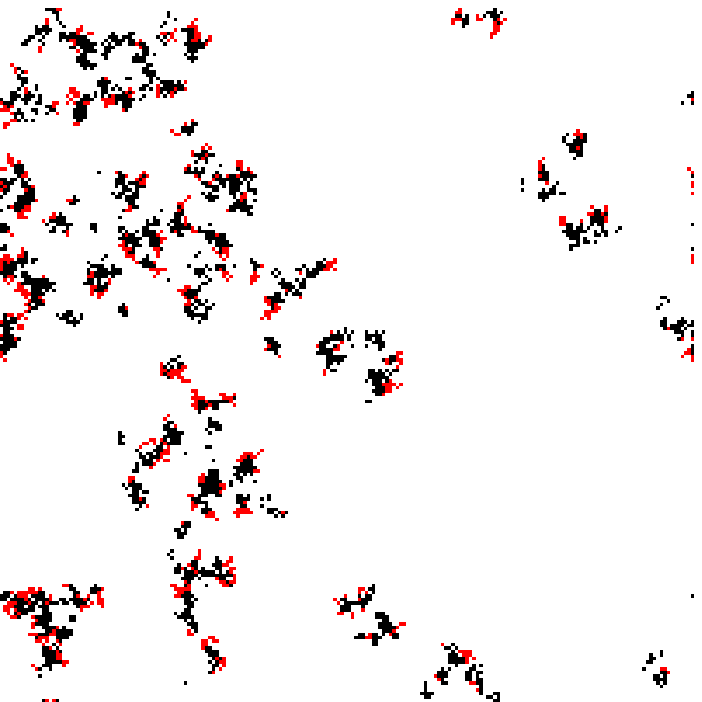}
(b)\includegraphics[width=4cm]{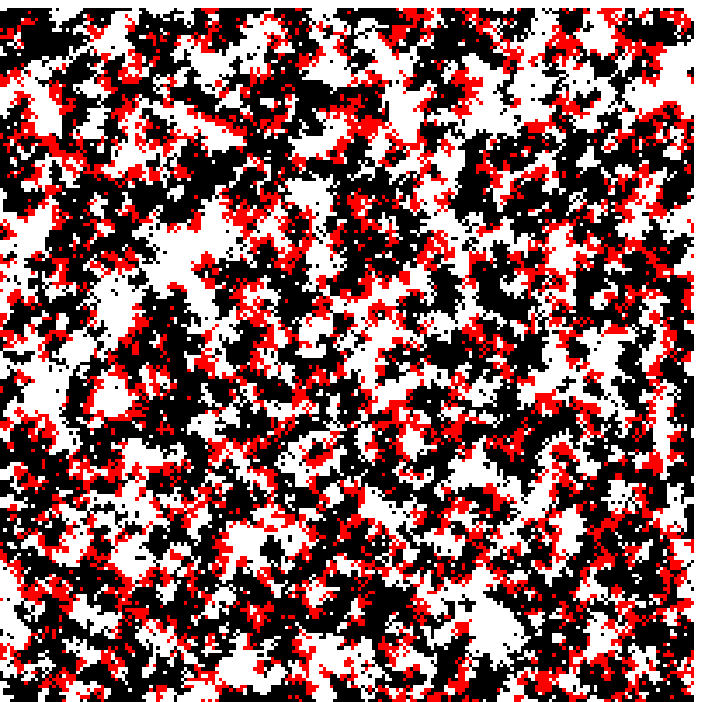}
(c)\includegraphics[width=4cm]{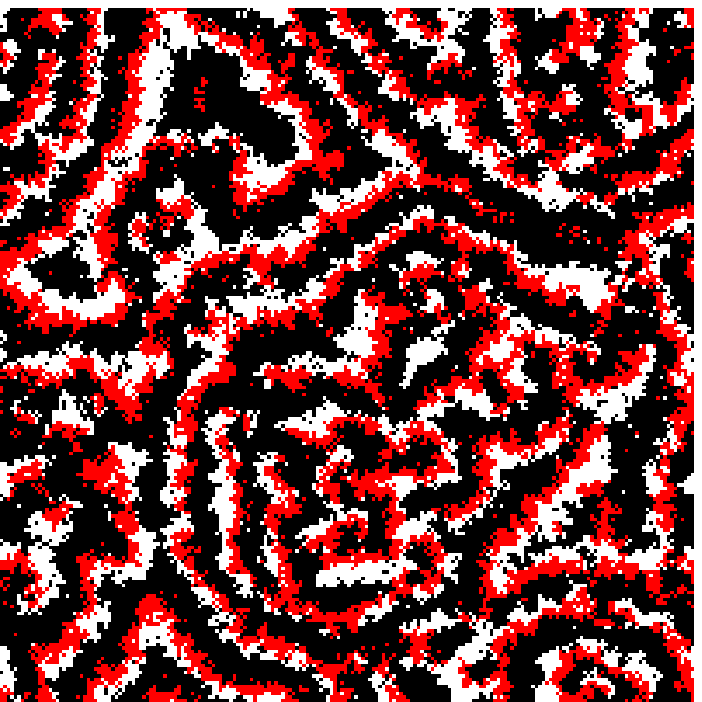}
\caption{\label{fig3} (Color online) Three types of spatial pattern
absent of mobility with $\tau_{I}=0.40$, $\beta_{min}=0.60$ and
$\beta_{max}=2.10$. Here three possible states are shown:
susceptible cells (white), infected (red), and resistant (black),
respectively. (a) Localized disease outbreaks are self-limiting in
size for $\beta=0.62$ ; (b) Turbulent waves for $\beta=1.20$; (c)
Stable spiral waves for $\beta=1.80$. Grid size for all panels is
$200\times 200$. Note that all the patterns arise from the same
initial condition. These spatial patterns can be produced by using
supporting online material (see the computer code for the cellular
automata at http://7y.nuc.edu.cn/jinzhen/English1.aspx). The random
number generator itself may be different if you are not using Linux,
but it does not make a qualitative difference. }
\end{figure}

The classical mean-field theory on the diseases spreading has shown
that the number of number of secondary cases due to a single
infected individual (referred $\mathcal{R}_0$ is maximized in the
model~\eqref{eq:mf} and therefore maximum transmission $\beta$ is
selected. Here, our results show that once the spatial structure is
included the $\mathcal{R}_0$ is no longer maximized and the
transmission rate is limited (see the labeled a in
Fig.~\ref{extinction}). We note that this results are consistent
with parasite evolution and extinction~\citep{Boots2003}. In
addition, in the spatial explicit model, the diseases can survive
for a lower infection rate than mean-field context.  This is a low
infection rate will tend to increase the local density of
susceptible individual around infectious. Spatial structure within
cluster of individual favors a lower infection rate. The other way
round, with higher infection rate will tend to infect all the
individuals in a cluster quickly. This in turn leads to  relatively
rapid local cluster extinction.

Moreover, we also find that, for the fixed infection period
$\tau_{I}$, the different moving spatial patterns emerge with
 infection rate $\beta$ increasing. Figure~\ref{fig3} shows the typical
 snapshots of the stable spatial patterns when the parameters are within
 domain of persistence II in Fig.~\ref{extinction}, in which the results indicate
 that the threshold values $\beta_{min}$ is equal to $0.60$ and $\beta_{max}$
 equal to $2.10$ for $\tau_{I}=0.40$ respectively. When the infection rate is
 low (but just above the critical value $\beta_{min}$),
we find that the susceptible and infectious individual coexist, and
self-organized spatial patterns with limited size of moving
clusters. It means that localized disease outbreaks are
self-limiting in size for this case (see Fig.~\ref{fig3}(a)). With
increasing infection rate $\beta$, these structures grow in size and
form pattern of moving turbulent/spirals (see Fig.~\ref{fig3}(b) and
(c)), and disappear for large enough $\beta$ (exceed the critical
value $\beta_{max}$).

\begin{figure}[htp]
\centering
\includegraphics[width=8cm]{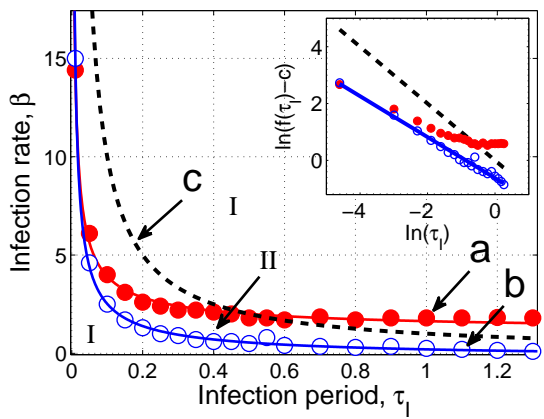}
\caption{\label{extinction} (Color online) Extinction and
persistence phase transition for ($\tau_{I}$,$\beta$)-parameters
space. The red circles and blue circles are estimating from the
simulation for $\beta_{max}$ and $\beta_{min}$ respectively. Red
line (marked by \textit{a}) is fitted by using  $f(x) = ax^b+c$
function and $a=0.5264 $ (0.4456, 0.6071),  $b=-0.7324$ ($-0.7647$,
$-0.7001$), and $c=1.008  $ (0.7124, 1.304).  Blue line (marked by
\textit{b}) is fitted by using $f(x) = ax^b+c$ function and $a=
0.6276 $ (0.4384,0.8168), $b=-0.6658$ ($-0.7287$, $-0.6029$), and
$c=-0.3256$ ($-0.4643$, $-0.1869$).  Dashed line (marked by c)
denotes the prediction obtained from mean-field theory on
$\mathcal{R}_0$. Inset shows the a nonzero $c$ (nonvanishing
infection rate) and zero $c$ cases respectively on the log-log plot.
}
\end{figure}

Bartlett observed in a series of
papers~\cite{Bartlett1956,Bartlett1957,Bartlett1960} that measles in
large cities had recurring outbreaks, while it went extinct in small
communities until reintroduced from external sources. This means
that the time to extinction is an increasing function of the
community size. However, the analysis by Nasell showed that
Bartlett's approximation of the extinction time was unsatisfactory
in an important part of parameter
space~\cite{Nasell1999,Nasell2005}. Our present
system~\eqref{eq:reaction} is a simple individual-based stochastic
SIRS model with diffusion
processes~\cite{Alonso2006,Simoes2008,Reichenbach2008,Richard2008}.
The goal of the analysis is to derive information about the
quasi-stationary distribution and the time to extinction. Pursuing
this goal leads to difficult mathematical problems. Exact solutions
cannot be found. A possible way to proceed is therefore to work with
approximations by simulation. Early efforts to derive approximations
of the expected time to extinction were shown by Nasell to lead to
large errors~\cite{Nasell1999}. In the present paper, a
approximation of the discrete state Markov chain~\eqref{eq:reaction}
is introduced. It led to a truncated normal distribution as an
approximation of the marginal distribution of infected individuals
in quasi-stationarity~\cite{Nasell2001,Nasell1999b,Nasell2005}. The
resulting approximation of the expected time to extinction was
rather coarse by a linear relation, but turned out to be an accepted
approximations for present analysis of extinction and persistence.

\begin{figure}[htp]
\centering
\includegraphics[width=8cm]{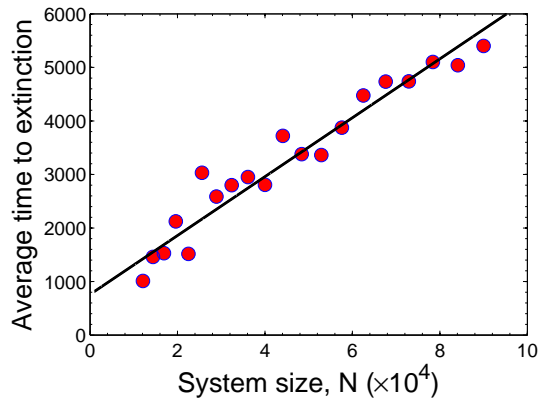}
\caption{\label{fig4} (Color online) Average time to extinction
$T_{ex}$ versus system size $N$ for $\tau_{I}=0.40$ and
$\beta=0.50$. (the least-squares linear fit of the data has a slope
of 0.055 and an intercept of 758.75 (see text for details).}
\end{figure}

The dependence of the average time ($T_{ex}$) to extinction  on the
system size of changes, $N$, is shown in Fig.~\ref{fig4}, which also
demonstrates the dependence of $T_{ex}$ on the size of the
populations when the parameters are within domain of extinction. As
seen, average extinction time increases linearly with system size
$N$ (we test the system size from 10,000 to 90,000). More
quantitatively, $T_{ex}$ is approximately proportional to $N$, which
is fitted the simulation results by using linear function, and then
obtain the relationship as $T_{ex}=0.055N+758.75$ are shown in
Fig.~\ref{fig4}. To ensure that the used averaging over just 100
independent runs yields good statistics, we present in
Fig.~\ref{fig4} the each data of $T_{ex}$ obtained 100 runnings for
$\tau_{I}=0.40$ and $\beta=0.50$ (these parameters corresponding to
extinction lower domain I in Fig.~\ref{extinction}.). It should be
noticed that previous phase transitions analysis is convincible
because third time to extinction was used to estimate phase
transitions of persistence and extinction. One can seen that 10
thousand time units are a long-term run for persistence on this
system. Here, it is instructive to note that our results show that
the persistence and extinction are independent on the population
size. The opposite view have been obtain from the well-mixed systems
in recently Ref.~\citep{claussen:058104}, in which they demonstrate
a critical population size above which coexistence is likely.

\subsection{Mobility promotes persistence on spatial epidemic model}

In fact, the individual always exhibits motion in the space. How
does individuals mobility, in addition to nonlinearity, affect the
system's behaviors (i.e., persistence and extinction, spatial
pattern, invasion speed and so on)? The insight of this important
issue can be gained from the cellular automata model. In particular,
one can compare the results of present paper with the previous
works~\citep{Ballegooijen12282004}. Two models are commonly used to
predict spatial spread of a disease. The first is the
distributed-contacts model, often described by a contact
distribution among stationary individuals. Distributed-contacts
models are particularly appropriate for the study of plant
disease~\citep{Bosch1988,Bosch1999}, but researchers are also using
the closely related framework of contact networks to study disease
transmission in human population~\citep{PhysRevE.66.016128,Rand2003}
or animal species. Notice that recently developing spatial moment
closure methods are also based on this
framework~\citep{Bolker1997,Dieckmann2000}. The second is the
distributed-infectives models, often described by the mobility of
infected individuals. This approach relies on the assumptions that
disease is transmitted through interactions between dispersing
individuals, and that infected individuals move in uncorrelated
random walks. Medlock and Kot developed a distributed-infectives
framework that uses a flexible kernel-based approach similar to that
employed in distributed-contact models~\cite{Medlock2003}. They
found that inappropriate application of either the
distributed-contact or distributed infectives approaches can
generate inaccurate projections of epidemic spread. Hence, here we
consider a scenario that the transmission process involves
components of both distributed-contacts and distributed
infectives--diffusion of infected individuals. For instance the
spread of rabies that foxes tend to be restricted to discrete home
ranges. In Central Europe, the home range size is about 4 $km^2$,
but this may differ considerably between areas (range 2.5-16
$km^2$)~\citep{Harris1988,Jeltsch1997}. Based on previous
subsection, now we investigate the individuals with a certain form
of mobility. Namely, at rate $\varepsilon$ all individuals can
exchange their position with a nearest neighbor. Each individual
randomly exchanges with its eight neighbor (North, West, South,
East, Northwest, Southwest, Southeast, and Northeast) at each time
step. This is reasonable for the realistic system. These exchange
processes lead to an effective mobility of the individuals.

In this subsection we analyze the dynamics of the system by
considering individual spatial diffusion process. In fact, the
susceptibles' mobility will also lead to the spread of the disease
since it changes the spatial distribution of the infected
individuals. In present paper, we consider the mobile individuals
including susceptible, infected, and resistant. Hereafter, we refer
the terms \emph{exchange}, \emph{mobility}, and \emph{diffusion} to
as the same meaning--the mobility of individuals within nearest
neighbors. Hence, we simply refer them as ``mobility". We refer to
the mobility as following
\begin{equation}\label{eq:diffusion1}
XI\stackrel{\varepsilon}{\longrightarrow} IX,
\end{equation}
where $X\in\{S, I, R\}$. It has to be noted this mobility is not
taken into account at the mean-field rate equation level, as well as
van Ballegooijen and Boerlijst's study~\cite{Ballegooijen12282004}. Denote $L$ the linear size of
the $d$-dimensional hypercubic lattice (i.e. the number of sites
along one edge), such that the total number of sites reads $N=L^d$.
Choosing the linear dimension of the lattice as the basic length
unit, the macroscopic diffusion constant $D$ of individuals stemming
from mobility processes reads $D=\varepsilon d^{-1}N^{-2/d}$ for a
continuum limit~\citep{reichenbach:238105,Reichenbach2008}. In this
limit, a description of the stochastic lattice system through
stochastic partial differential equations (SPDE) becomes feasible.
But here we study the system's behavior in the approach of the
cellular automata model.

\begin{figure}[htp]
\centering
\includegraphics[width=7cm]{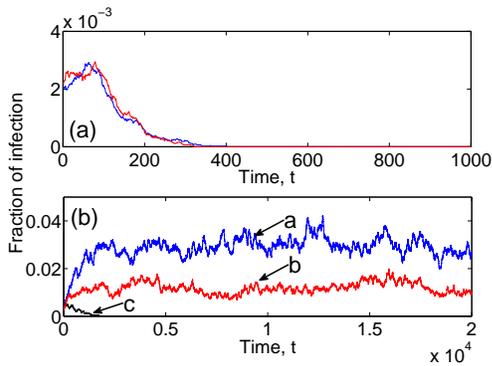}
\caption{\label{fig-diffusion1} (Color online) The time-series of
infected fraction before and after the mobility is turned on with
$\beta=0.42$ and $\tau_{I}=0.40$. (a) Two independents run before
mobility rate is turned on. (b) For different mobility rate
$\varepsilon$ is turned on, line \textit{a}: $\varepsilon=10.0$,
line \textit{b}: $\varepsilon=8.0$, and line \textit{c}:
$\varepsilon=5.0$.}
\end{figure}

\begin{figure}[htp]
\centering
\includegraphics[width=7cm]{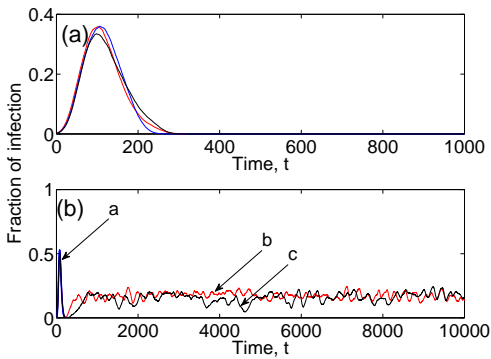}
\caption{\label{fig-diffusion2} (Color online) The time-series of
infected fraction before and after the mobility is turned on with
$\beta=2.50$ and $\tau_{I}=0.40$. (a) Three independents run before
mobility rate is turned on. (b) For different mobility rate
$\varepsilon$ is turned on, line \textit{a}: $\varepsilon=10.0$,
line \textit{b}: $\varepsilon=8.0$, and line \textit{c}:
$\varepsilon=5.0$.}
\end{figure}

We performed cellular automata simulations of
Eq.~\eqref{eq:reaction}  with the mobility processes
Eq.~\eqref{eq:diffusion1}. The space and time steps were chosen the
same as Subsection~\ref{section:short}. In our simulation, the
relationship between mobility rate and the probability of mobility
apply the Gillespie algorithm~\citep{Gillespie,gillespie:297}.
 We wish to examine how the movements of individuals affect the persistence
 and extinction of
 epidemic in the spatially structured population. In particular, we will examine
 the role of mobility's strength $\varepsilon$ when the parameters are less than the
 critical values  $\beta_{min}$. First, we investigate the time series of
 the infected fraction. We
 have kept the infection period $\tau_{I}$ fixed at a value $\tau_{I}=0.40$,
 and systematically varied the mobility rate $\varepsilon$. In
 Figs.~\ref{fig-diffusion1} and \ref{fig-diffusion2}, we show typical
 long-time series of the infections for various values of the mobility rate
 when the parameter $\beta=0.42$ (below the critical values $\beta_{min}\approx 0.6$)
 and $\beta=2.50$ (above the critical values $\beta_{max}\approx 2.10$), respectively.
 Fig.~\ref{fig-diffusion1}(a) and (b) show  a family time series
 of infected fraction
  before and after the mobility rate is turned
  on respectively, different $\varepsilon$ values marked by
  \textit{a}, \textit{b}, and \textit{c} in Fig.~\ref{fig-diffusion1}(b).
We observe in Figs.~\ref{fig-diffusion1} that the mobility raises
the possibility of shift from extinction to persistence state in the
spatial epidemics model. Furthermore, this persistence with
irregular fluctuation is around a positive equilibrium. It is worth
noting that the large mobility rate $\varepsilon$ can promote the
persistence of spatial epidemics in despite of the parameter $\beta$
below the minimal critical value $\beta_{min}$. Our simulations
indicate that the similar results hold when the parameter above the
maximal critical value $\beta_{max}$, but in which the persistence
corresponds to low mobility rate $\varepsilon$ (see
Fig.~\ref{fig-diffusion2}).

\begin{figure}[htp]
\centering
\includegraphics[width=7cm]{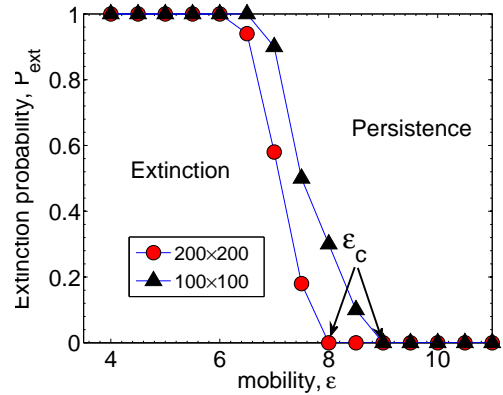}
\caption{\label{fig-extinction3} (Color online) The critical
mobility $\varepsilon_{c}$ with different system sizes. Mobility
above the value $\varepsilon_{c}$ induces persistence; while it is
extinction below that threshold. The data obtained from lattice
simulations of the system after long temporal development (that is,
at time $t\propto N$) and for different values of $\varepsilon$ with
$\beta=0.42$ and $\tau_{I}=0.40$. Here, we have considered the
extinction probability $P_{ext}$ starting with randomly distributed
individuals on a $200\times200$ square lattices and $100\times100$
square lattices, respectively. }
\end{figure}

Next, we examine the effect of mobility on the regions where the
epidemic will extinct when the mobility rate is varied. By using the
method developed in literature~\citep{reichenbach2007nature}, we
know that this scenarios can be distinguished by computing the
probability $P_{ext}$ when the epidemic has gone extinct after a
waiting time $t\propto N$. For illustration, we have considered the
fixed parameters $\beta=0.42$ and $\tau_{I}=0.40$. In
Fig.~\ref{fig-extinction3}, we obtain the dependence of $P_{ext}$ on
the mobility rate $\varepsilon$. With increasing mobility rate, the
extinction probability with a sharpened transition emerges at a
critical value $\varepsilon_{c}=8.0\pm0.05$ for the entire
$200\times200$ lattice area explored in 10000 time-unit. One could
see that, above $\varepsilon_{c}$, the extinction probability
$P_{ext}$ tends to zero as the mobility rate increase, and the
epidemic is stable persistence (implying super-persistent
transients~\citep{Hastings:y2004:0169-5347:39}). On the other hand,
below the critical mobility rate $\varepsilon_{c}$, the extinction
probability approaches 1 for small mobility rate, and the epidemics
occurs is unstable. One of our central results is that we have
identified a mobility threshold for persistence of epidemics.
Without loss of generality, we also have tested the $P_{ext}$  data
for the other parameters values $\tau_{I}$ and $\beta$, the same
sharpened transitions have been observed from the lattice
simulations.

There exists a critical value $\varepsilon_{c}$ such that a high
mobility $\varepsilon>\varepsilon_{c}$ changes the extinction phase
to persistence phase, while $\varepsilon<\varepsilon_{c}$ remains
extinction, leaving a uniform state with only susceptible in entire
space. It is worth noting that our above results arise from the
domain I of extinction phase in Fig.~\ref{extinction} where
$\beta<\beta_{min}$ and absence of mobility. Therefore, our results
also indicate that the stronger mobility can increase the parameter
region where the epidemic persistence occurs. This may be completely
explained from the local and global infection~\citep{Boots1999}. The
higher mobility leads to a increasing of the global contact among
individuals in the lattices.

\begin{figure}[htp]
\centering (a)\includegraphics[width=4cm]{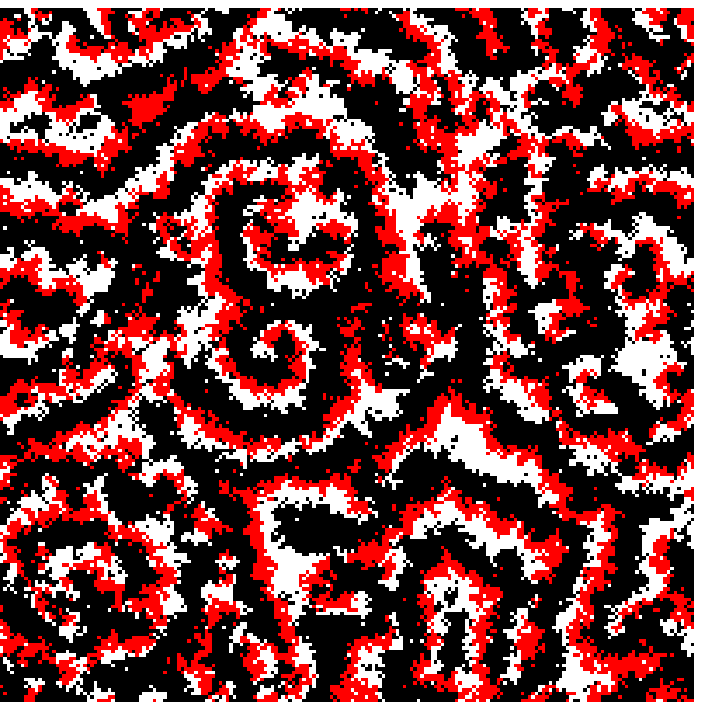}
(b)\includegraphics[width=4cm]{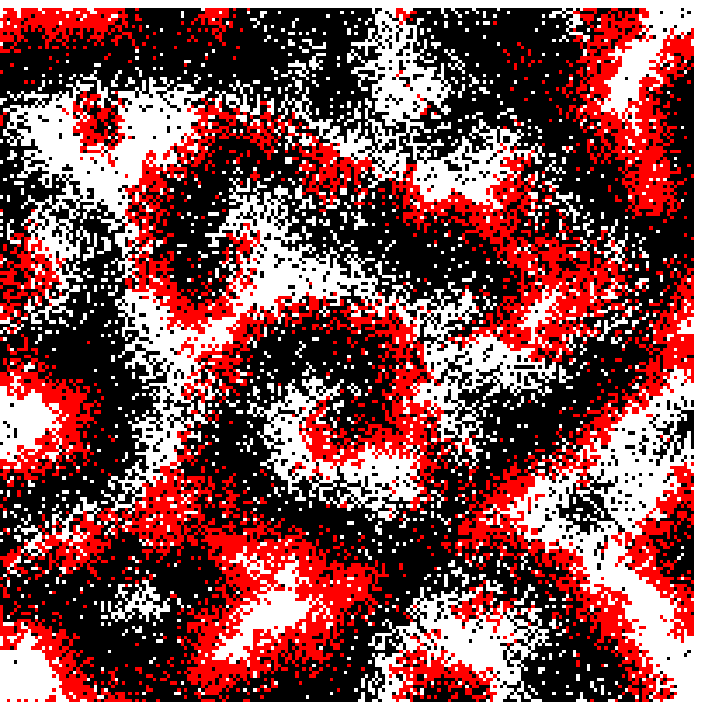}
\caption{\label{fsm} (Color online) The effect of mobility on
spiralling patterns. We show snapshots obtained from lattice
simulation of typical states of the system after long temporal
development and for different values of $\varepsilon$ starting with
randomly distributed individuals on a square lattice. (a)
$\varepsilon=0.01$; (b) $\varepsilon=8.05$. Parameters are
$\tau_{I}=0.40$ and $\beta=1.80$.}
\end{figure}

\begin{figure}[htp]
\centering (a)\includegraphics[width=4cm]{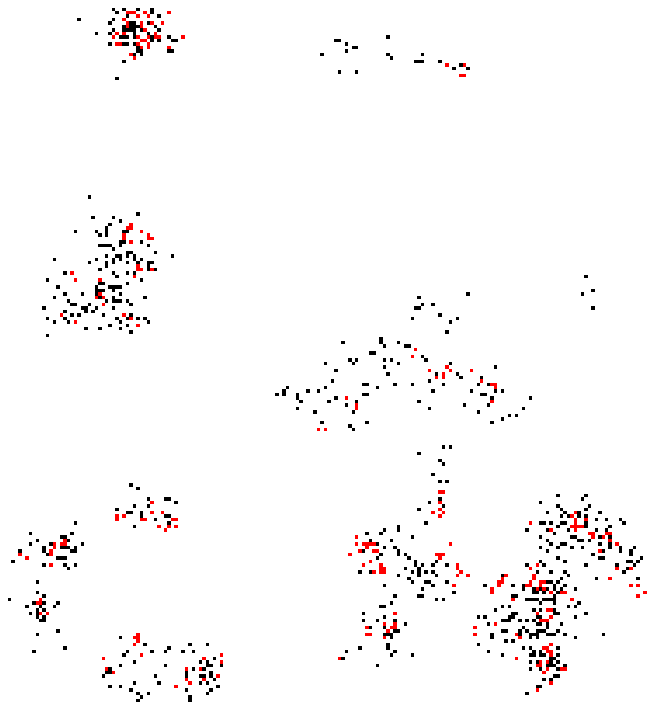}
(b)\includegraphics[width=4cm]{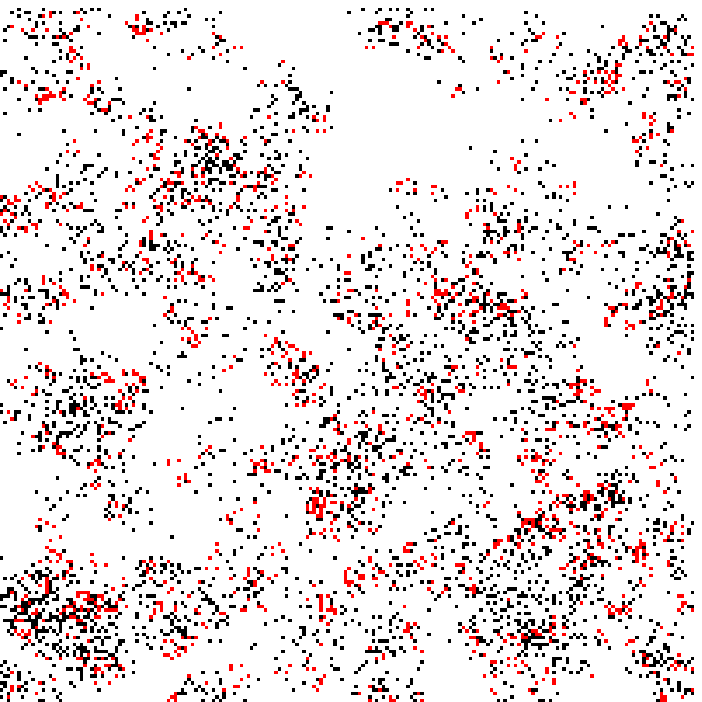}
\caption{\label{fig-diffusion3} (Color online) Spatial pattern in
the lattice simulation for mobility above the value
$\varepsilon_{c}$ induces persistence with $\beta=0.42$ and
$\tau_{I}=0.40$. Here, we have considered the spatial pattern
starting with randomly distributed individuals on a square lattice.
(a) $\varepsilon=8.0$; (b) $\varepsilon=12.0$.}
\end{figure}

Reichenbach \textit{et al}  had assessed the effects of mobility on
the spatial pattern of rock-paper-scissors
games~\cite{reichenbach2007nature}. Results demonstrated that a
critical influence of mobility on species diversity: when mobility
exceeds a certain value, biodiversity is jeopardized and lost. In
contrast, below this critical threshold all subpopulations coexist
and an entanglement of traveling spiral waves forms in the course of
time. In our study, we incorporate the scenarios outlined in
Refs.~\cite{reichenbach2007nature,Czaran2002} and also investigate a
spatial pattern induced by mobility when the parameters are within
domain II of Fig.~\ref{extinction}. We perform extensive computer
simulations of the stochastic system (see Eqs.~\eqref{eq:reaction}
and \eqref{eq:diffusion1}) and typical snapshots of the steady
states are shown in Fig.~\ref{fsm}. With increasing mobility
$\varepsilon$, these moving spiral structure grow in size and
saturate at a certain value, but they don't disappear with respect
time. This is different from rock-paper-scissors games where the
spiral wave will disappear for large
mobility~\cite{reichenbach2007nature}. As shown in Fig.~\ref{fsm},
the spirals' wavelength $\lambda$ rises with the individuals'
mobility increased. When the parameters are within domain I, we have
not observed the spatial pattern emergence for varied mobility rate
$\varepsilon$. However, we find that the high mobility lead to
entire disease outbreaks instead of localized disease outbreaks (see
Fig.~\ref{fig-diffusion3}).

From the above observations we know that the mobility of individuals
can
 dramatically affect the epidemics extinction/persistence and spatial pattern.
 We need  to point out here that, although our results in this study come
 from cellular
 automata's simulation, it is also useful for assessing the dynamics of
 deterministic systems, such as stochastic PDEs, spatial correlation models,
 ordinary difference equation, and so forth~\citep[see][]{reichenbach2007nature,
 reichenbach:238105, Reichenbach2008,Keeling1999,Ebert2000}. By modeling the
 interaction among individuals, we are able to understand the role of spatial
 mixed (cause by mobility) in invasion dynamics without the need for
complex mathematical methods. Whether an organism can successfully
invade and persist in the long-term is dependent on many factors.
Here, we have identified a mobility factor for epidemic persistence
and invasion, which is distinguished from the nonlinear dynamics of
the epidemic models.

\subsection{Spatio-temporal synchronous dynamics of spatial epidemics}

\begin{figure}[htp]
\centering
\includegraphics[width=7cm]{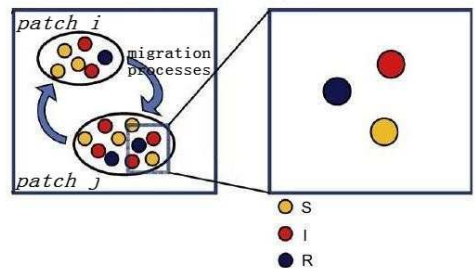}
\caption{\label{figmulti} (Color online) Typical spatial structured
model of the two-patches spread of epidemics (schematic). The two
patches $i$ and $j$ are connected by migration processes. Each patch
contains a population of individuals who are characterized with
respect to their stage of the disease (e.g. susceptible, infected,
resistant), and identified with a different color in the picture. }
\end{figure}

As individuals travel around the world for human, as well as migrate
for animals, the disease may spread from one place to another. The
spatial spread of epidemics has been much studied, particularly with
respect to pandemic invasion
waves~\citep{CecileViboud04212006,Yingcun2004,Earn1998,Grais2003,
Hufnagel2004,Grassly2005}. Simulation models incorporating
transportation have generated important insights into the spread of
epidemics. However, the key underlying relationship between human
movement and disease spread has not been verified across wide
spatial scales. To quantify the traveling behavior of individuals,
we consider the linked dynamics of two host communities each of size
$N$, where individuals possess mobility in each patch. Spatial
coupling is the critical parameter determining phase coherence and
spatial synchrony~\citep{Grenfell2001} and the persistence of
host-pathogen systems~\citep{Swinton1998,Grenfell1998}. For
simplicity, we assume that $N$ is constant through time for each
patch. The typical schematic for two patches is shown in
Fig.~\ref{figmulti}.

\begin{figure}[htp]
\centering
\includegraphics[width=7cm]{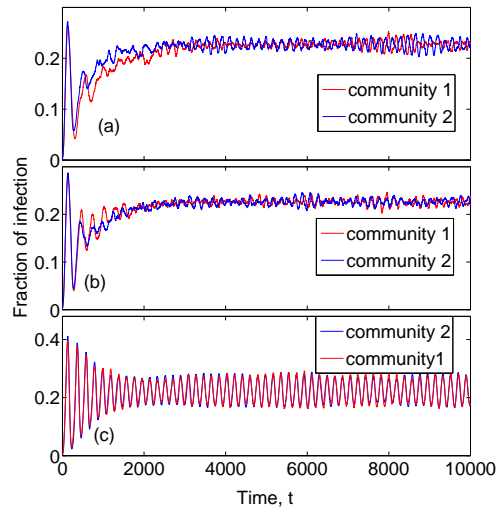}
\caption{\label{fig-patch} (Color online) The time series of
infected fraction in two coupled communities with $\varepsilon=0$,
$\beta=1.80$, and $\tau_{I}=0.40$. (a) With weak coupling,
$LR=0.001$, rapidly anti-phase synchronization predominates; (b)
With middle coupling, $LR=0.01$, remaining anti-phase
synchronization predominates; (c) With strong coupling, $LR=0.1$,
epidemic peaks, and the troughs between them, rapidly synchronize
in-phase.}
\end{figure}

\begin{figure}[htp]
\centering
\includegraphics[width=7cm]{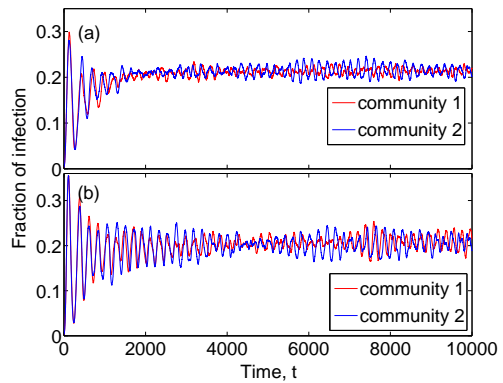}
\caption{\label{fig-patch2} (Color online) The time series of
infected fraction in two coupled communities with different mobility
rate inside each patch. (a) Low mobility rate, $\varepsilon=0.5$,
anti-phase synchronization predominates; (b) With hight mobility
rate, $\varepsilon=4.5$, remaining anti-phase synchronization
predominates. The other parameters are same as in
Fig.~\ref{fig-patch}(b).}
\end{figure}

Consider now, two patches with sizes $200\times200$ cells each
other, each representing a community or suburb, and suppose a small
number of infective individuals in each community at initial time.
The individual can pass from one community to the other. This
migration or the individual traveling is randomly chosen for some
given individuals moving from a community to another community in
the large metapopulation systems. The strength of coupling between
the two communities is measured by
\begin{equation}
LR=\texttt{no. of migration}/\texttt{size of community}
\end{equation}
with respect to unit time. Although the approach is only a rough
approximation of the dynamic nature of transportation between two
communities, it nevertheless capture some of the essential
elements~\citep{Daihai2003,Colizza2008}. Figure~\ref{fig-patch}
shows a simulation results with different coupling strength between
two communities, in which the mobility within patch is ignored. In
this case, the infectives in each community are seen to irregular
oscillate and local anti-phase for weak coupling strength (see
Fig.~\ref{fig-patch}(a) and (b)). Upon increasing the coupling
strength between the communities, the suburbs show in-phase
synchronization~\footnote{Note that here the synchronization or
asynchronization refers to oscillate with Uniform Phase evolution,
yet has Chaotic Amplitudes(UPCA). In addition, the synchronization
can be quantified by calculating the phase of each time-series of
infectives and plotting the phase difference between the two
suburbs~\citep{PhysRevLett.76.1804, Blasius2000,Grenfell2001}.} (see
Fig.~\ref{fig-patch}(c)). Notice that the similar results have been
obtained by using SIRS network
model~\citep{Daihai2003,PhysRevLett.86.2909}. As previous
subsection's results show that, at the single patch level, the
epidemic behavior on the global~\footnote{Here, the global refers to
as the single patch, rather than the metapopulation system.} scale
is also determined by the mobility of individuals. In particular,
the effects due to the finite size of communities and the stochastic
nature of the mobility might have a crucial role in the problem of
resurgent epidemics, extinction and eradication~\citep[also
see][]{Ball1997,Cross2005,Cross2007,Colizza2008}. Therefore it is
important to consider the effect of mobility rate on synchronization
for the metapopulation system. In Fig.~\ref{fig-patch2}, we report
the effect of different mobility rate on the time-series of infected
fraction by the lattice simulations. Figure.~\ref{fig-patch2}
provides a clear evidence of the spatio-temporal synchrony in
metapopulation system independent of the self-mobility, but it
depends on the coupling strength among patches. One can draw a
conclusion that mobility within each patch does not play a
signification role, and spatial structure within patches can be
ignored when investigate the coupling patches.

One of our main goals is to study the manner in which two or more
communities or suburbs synchronize in time, as recurring waves of
infection sweep through their respective populations by means of
cellular automata simulation. This is a fascinating outcome of
spatial dynamics whereby a few migrating infective individuals have
the potential to spread the epidemic from one community to the
other, giving rise to synchrony in the long term. For instance, the
time-series of measles infections in various cities of England
(1944--1958) are reported in literatures~\citep{Grenfell2001,
Daihai2003,Bharti2008}. The Birmingham and Newcastle appear to
synchronize in-phase together while Cambridge  and Norwich are
clearly out of phase by 180 degree (see Fig.~\ref{fig-measles1} in
Appendix and Fig.~2 in literature~\citep{Grenfell2001,Bharti2008}).
Here, we simulate a simple coupled spatial community models to gain
insights about the coupling strength. These results provide useful
insights for the basic theoretical understanding of mechanistic
epidemic models in complex multi-patches environments, which can
then be used to build more realistic data-driven large-scale
computational approaches for real case scenarios and spatially
targeted control measures. For instance, the spread of influenza
shows spatial synchrony in 49 states of United
Sates~\citep{CecileViboud04212006}.


\section{Conclusions and discussion}\label{sec:discussion}

As a summary, from the simulations of individuals' spreading
processes among lattices with randomly distributed in the space, we
can identify different effects of parameters describing the spatial
structure and the temporal developments of disease for individuals,
such as $\tau_{I}$, $\beta$, $\varepsilon$, $LR$ and $N$, on the
dynamical behavior--persistence, extinction and synchrony. We change
every parameter which we consider can affect the spreading process.
Although the short-term behavior (persistence and extinction) may
depend on system size $N$ and details of spatial distribution, the
long-term behavior mainly depends on three parameters: the infection
period $\tau_{I}$, infection rate $\beta$, and mobility
$\varepsilon$ for single community. The two former parameters
determine what types spatial pattern emerge, and the last
dramatically increase the parameters region where the epidemics
pandemic occurs in the space. Moreover, we investigate the coupling
between multi-community, in which the two former parameter
$\tau_{I}$ and $\beta$ reflect the nature of the developing and
pandemic period of disease for individuals, the coupling strength
$LR$ is a parameter related to the population density, the
frequencies of people contacts, and the extent of people traveling
between different places or cities~\citep{Grenfell1998}. These
results may be helpful for the analysis of the spreading processes
of diseases in space.

The possibility of spiral waves, self-organized spatial pattern,
spotted pattern, traveling wave as well as trade-offs in spatial
epidemics and host-parasitoid by local dispersal abilities has long
been
recognized~\citep{Comins1996,Hassell1994,Ballegooijen12282004,liu:031110,
1742-5468-2007-05-P05002,1742-5468-2007-11-P11011,CecileViboud04212006,Grenfell2001}.
What is new here is that the trade-off between persistence and
extinction have been understood in the spatial structure epidemics
model. The discrete character of the individuals involved in the
reactions~\eqref{eq:reaction} and the mobility
processes~\eqref{eq:diffusion1} are responsible for intrinsic
stochasticity arising in the system. Individuals' mobility as well
as intrinsic noise have crucial influence on the self-formation of
spatial patterns. The analytical expressions for the spirals'
wavelength as function of mobility can be determined by means of a
complex Ginzburg-Landau equation (CGLE) obtained by recasting the
PDE derived from interacting particle
approach~\citep{Reichenbach2008,reichenbach2007nature}. Hence, to
qualitatively explain these findings, a profound understanding is
still desirable and could motivate further investigations.

\section*{Acknowledgments}
We are grateful for the constructive suggestions of the two
anonymous referees on our original manuscript. We thank Bai-Lian Li
for improving the English on the original manuscript.

 This work was supported by the National Natural Science
Foundation of China under Grant No. 60771026, Program for New
Century Excellent Talents in University (NCET050271), and the
Natural Science Foundation of Shan'xi Province Grant No. 2007021006.

\appendix
\section{}

We give here the time series of weekly measles case reports for
Birmingham and Newcastle between 1948 and 1969, which exhibits
in-phase pattern.
\begin{figure}[htp]
\centering
\includegraphics[width=7cm]{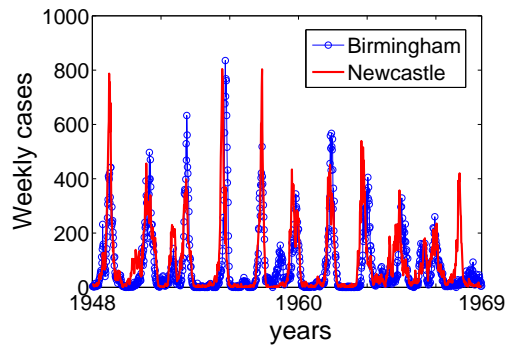}
\caption{\label{fig-measles1} (Color online) The time series of
weekly measles case reports for Birmingham and Newcastle between
1948 and 1969. Data made available by Professor Benjamin Bolker
(www.zoo.ufl.edu/bolker/measdata.html).}
\end{figure}

\newpage



\end{document}